\documentclass[11pt]{article}
\usepackage{jheppub}
\usepackage[T1]{fontenc}
\usepackage[latin9]{inputenc}
\usepackage{amsmath}
\usepackage{amssymb}
\usepackage{setspace}
\usepackage{float}\usepackage{amstext}
\makeatother

\begin{document}

\title{On the Vacuum Structure of the 3-2 Model}

\author{Tomer Shacham}

\affiliation{Racah Institute of Physics, \\
 The Hebrew University, \\
 Jerusalem 91904, Israel} 

\emailAdd{tomer.shacham@phys.huji.ac.il}

\abstract{ The 3-2 model of dynamical supersymmetry breaking is
revisited, with some incidentally new observations on the vacuum structure. 
Extra matter is then added, and the vacuum structure is further
studied. The parametric dependence of the location of the vacuum provides a consistency check of Seiberg duality.}
   

\maketitle
\flushbottom
\section{Introduction}

The first step towards understanding the spectrum and dynamics of a quantum field theory is to study its vacuum structure. If the vacuum is strongly coupled, there is currently no description of its structure. However, in certain cases, $\mathcal{N}=1$ supersymmetry (SUSY) allows an insight to the low energy physics given by Seiberg duality \cite{SeibergDuality}. 

Since the duality still lacks a rigorous proof, consistency checks are of theoretical value.\footnote{Available evidence supporting the duality includes the following: \newline The dual theories have the same non-anomalous global symmetries; anomaly matching conditions for these symmetries are satisfied. If a SUSY vacuum exists, the dimension of the moduli space is the same in both theories. Other nontrivial checks include corresponding dualities in $\mathcal{N}=2$ theories \cite{SeibergArgyresPlesser}, matching of the superconformal index \cite{Romelsberger,DolanOsborn} and a brane construction in type IIA string theory \cite{BraneReview,BranesPictureOfDuality}.} Following the footsteps of \cite{ZoharThomasSudanoDterms}, a new non-holomorphic consistency check is presented here. The main idea is that the electric and magnetic theories cannot both be valid in overlapping regions of parameter space, as it is inconceivable that there are two different weakly coupled descriptions of the same physics.

A necessary condition for a perturbative description to be valid is that the vacuum is in fact calculable. In the context of asymptotically free gauge theories, the vacuum is considered calculable if the entire gauge group is Higgsed by expectation values parametrically larger than the scale of strong coupling. When this happens, quantum corrections are relatively small and a classical analysis of the vacuum structure is justified. 

Regarding Seiberg duals, the situation is different. As this is a strong-weak duality, the low energy descriptions of some asymptotically free theories are weakly coupled. The classical analysis of the vacuum structure is then reliable for small values of the charged fields.

If SUSY is a symmetry of nature, it must be broken.  For reasons of naturalness, it should be broken dynamically. The 3-2 model \cite{3-2Model} (recently reviewed in \cite{DineMasonReview}) is an example of dynamical SUSY breaking. Due to its simplicity, or perhaps the resemblance to the Standard Model, it has been extensively used as a model building tool \cite{SemiDirect,Raxion}. 

In this note, the vacuum structure of the 3-2 model is revisited, and some new observations are presented. Mainly, a vacuum is found in a previously unexplored region of parameter space. In order for a dual description to exist, we add two massive vector-like pairs of $SU(3)$. The vacuum is then located in both the electric and magnetic theories; the two regions of parameter space where the descriptions are valid indeed do not overlap. It should be noted that since the 3-2 model does not have a SUSY vacuum for any choice of parameters, this work is independent of previous~checks.
 
One aspect of the analysis is particularly interesting: the requirement that both descriptions have the same global symmetry is met in a non-trivial way. If the superpotential coupling is the smallest parameter, a single  vacuum exists in the electric theory. In the magnetic theory, however, there seem to be two distinct parametrizations of vacua but no residual gauge symmetry left to connect them. As it turns out, these two vacua coincide at the 2-loop~order.

\section{The 3-2 model\label{sec:The-3-2-Model}}

\subsection{A brief review}
The 3-2 model is an $\mathcal{N}=1$ supersymmetric gauge theory with matter content  
\begin{center}
\begin{tabular}{c|c|c}
 & $\left[SU(3)_{\textrm{C}}\right]$  & $\left[SU(2)_{\textrm{L}}\right]$\tabularnewline
\hline 
$Q_{A}^{a}$  & $\Box$  & $\Box$\tabularnewline
$\bar{Q}_{a}^{\alpha}$  & $\bar{\Box}$  & 1\tabularnewline
$L_{A}$  & $1$  & $\Box$\tabularnewline
\end{tabular}
\par\end{center}
where $\alpha=1,2$ is a flavor index. In the limit $\Lambda_{3}\gg\Lambda_{2}$,
the vacuum structure can be understood as follows. Consider
 $SU(2)_{\textrm{L}}$ to be a global symmetry.  A non-perturbative superpotential\begin{equation}
\mathcal{W}_{\mathrm{ADS}}=\frac{2\Lambda^{7}}{\mathrm{det}\left(\bar{Q}Q\right)}
\label{Wnp}
\end{equation}
pushes the fields away from the origin of field space. It is therefore natural to describe the vacuum in terms of the microscopic fields rather than the gauge invariants, to allow for a canonical K\"ahler potential at large fields. In this regime, $\mathcal{W}_{\mathrm{ADS}}$ is generated by an $SU(3)$ instanton. 
In attempt to stabilize this runaway, one may add interactions at tree level:\footnote{ This superpotential is generic; a term $\tilde{\lambda}\bar{Q}^{2}QL$ could be removed by a redefinition of $\bar{Q}$.}
\begin{equation}
\mathcal{W}_{\mathrm{tree}}=\lambda \bar{Q}^{1}QL.
\end{equation}

To a first approximation, the vacuum can be found on the $SU(3)_{\textrm{C}}$ D-flat directions as this gauge coupling is the largest parameter in
the theory. For convenience, we define $\Delta\equiv\left(g_{2}/ \lambda \right)^{2}$, 
where $g_2$ is the $SU(2)$ coupling. In the limit $\Delta\gg1$, the vacuum is found at the minimum of
$\mathcal{F}\big|_{\mathcal{D}=0}$. The VEVs of all fields scale like $\lambda^{-1/7}\Lambda$
and the vacuum energy scales like $\lambda^{10/7}\Lambda^{4}$. The
smallness of $\lambda$ ensures both that the gauge group is higgsed at scales much higher than $\Lambda$ and small SUSY breaking.

Recently \cite{ZoharThomasSudanoDterms}, the vacuum structure was found in
the opposite limit. When $\Delta\ll1$, it can be understood
as follows. Consider $SU(2)_{\textrm{L}}$ a global symmetry and assume that
 $L_{2}$ is stabilized far from the origin. This generates
a large mass $\left|\lambda\, L_{2}\right|$ for $Q_{1},\bar{Q}^{1}$
and these fields decouple. For large values of the remaining flavor
$Q_{2},\bar{Q}^{2}$, the gauge group is spontaneously broken to
$SU(2)_{\textrm{C}}$. Moduli space dependence of the gaugino condensate results in a runaway potential; this direction is lifted by weakly gauging $SU(2)_{\textrm{L}}$. The vacuum sits at $L_{2}\sim Q_{2}\sim\Lambda\left(\lambda\,\Delta\right)^{-1/7}$
and the vacuum energy scales like $\Lambda^{4}\lambda^{10/7}\Delta^{3/7}$.
The smallness of $\Delta$ justifies the integration out of the first
flavor - the condition is $\lambda\, L_{2}\gg\Lambda$ (equivalent to
$g_{2}\ll\lambda^{4}$).

\subsection{Some new observations}

We now show that the non-perturbative superpotential (\ref{Wnp}) is generated by an instanton even in the regime $\Delta\ll1$ where the D-term and F-term are comparable. In practice, we find the vacuum. As before, we start by considering $SU(2)_{\textrm{L}}$ to be global. 

Modulo gauge transformations, $Q_{A}^{a}$ has the form $\Bigg(\begin{array}{ccc}
a & 0 & 0\\
0 & b & 0
\end{array}\Bigg)^{T}$ with $a,b\in\mathbb{R}$. Since $\mathcal{W}_{\mathrm{tree}}$ breaks the flavor symmetry in the $\bar{Q}$ sector, the most general form of $\bar{Q}$ which respects the $SU(3)$ D-flat equations is $Q^{\dagger}\,\mathcal{G}$ with $\mathcal{G}\in SU(2)$.\footnote{For 2 flavors of $SU(3)$, the D-flat equations are $Q_{a}^{A\dagger}Q_{A}^{b}-\bar{Q}_{a}^{\alpha}\bar{Q}_{\alpha}^{\dagger b}=0.$}
 A convenient parameterization~is 
\begin{equation}
\mathcal{G}=\Bigg(\begin{array}{cc}
e^{i\phi}\cos\theta & e^{i\chi}\sin\theta\\
-e^{-i\chi}\sin\theta & e^{-i\phi}\cos\theta
\end{array}\Bigg).
\end{equation}
After scaling all fields in {}``units'' of $\lambda^{-1/7}\Lambda$, we find that the
F-term is $\mathcal{F\,}\Lambda^{4}\lambda^{10/7}$ where 
\begin{eqnarray}
\mathcal{F} & \equiv & \frac{4}{a^{4}b^{6}}+b^{2}l_{1}^{2}+\cos^{2}\theta\left[4\frac{\left(a^{2}+b^{2}\right)}{a^{6}b^{6}}+a^{2}\left(a^{2}+l_{1}^{2}+l_{2}^{2}\right)\right]-\frac{8}{a^{2}b^{2}}l_{1}\cos\left(\theta_{1}-\chi\right)\sin\theta+\nonumber \\
 & + & \left(a,b,l_{1},l_{2},\theta_{1}-\chi,\theta\right)\rightarrow\left(b,a,l_{2},l_{1},\theta_{2}+\phi,\theta+\pi/2\right)
\label{Fterm}
\end{eqnarray}
and $L_{A}$ is parametrized as $\left(l_{1}e^{i\theta_{1}}\,,\, l_{2}e^{i\theta_{2}}\right)$.\footnote{The substitution in the second line of (\ref{Fterm}) applies to all of the terms which appear in the first line.}
Note that $\mathcal{F}$ depends only on the combinations $\left(\theta_{2}+\phi\right)$
and $\left(\theta_{1}-\chi\right)$. This happens because the scalar potential
has additional Abelian global symmetries; these can be used to set
$\phi=\chi=0$. 

One can check that $\mathcal{F}$ has remaining runaways.
These are lifted by the weak gauging of $SU(2)_{\textrm{L}}$,\footnote{ $SU(2)_{\textrm{L}}$ generators are normalized such that $\textrm{Tr}T^{a}T^{b}=\frac{1}{2}\delta^{ab}$.} which generates a D-term $\mathcal{D\,}\Delta\Lambda^{4}\lambda^{10/7}$
where
\begin{equation}
\mathcal{D}\equiv\frac{1}{8}\left[4l_{1}^{2}l_{2}^{2}+\left(a^{2}-b^{2}+l_{1}^{2}-l_{2}^{2}\right)^{2}\right].
\end{equation}
 The full scalar potential, $\Lambda^{4}\lambda^{10/7}\left(\mathcal{F}+\Delta\mathcal{D}\right)$, is minimized at $\theta=0$ for all values of $\Delta$.\footnote{A field configuration corresponding to $\theta=\pi/2$ is connected by a gauge transformation.}

\begin{figure}[h]
\centerline{
\epsfig{file= 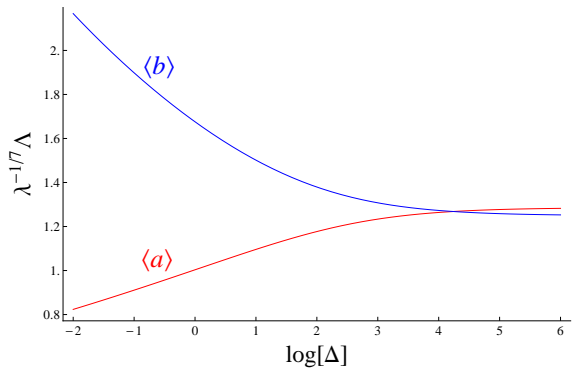, angle=0,width=7.5cm}}
\caption{
\footnotesize{The location of the vacuum for $\Delta\sim1$. }}
\label{abVacuum}
\end{figure}

In the absence of a D-term, $\mathcal{F}$~has a runaway to 
\begin{equation}
\left\langle L\right\rangle =\left(0,2a^{-4}b^{-2}\right),\qquad\left\langle b\right\rangle \rightarrow\infty.\label{eq:VEVL}
\end{equation}
For $\Delta\ll1$, the vacuum is found at the minimum of $\left(\mathcal{F}+\Delta\mathcal{D}\right)\big|_{\textrm{runaway }\mathcal{F}}$ .
 On the runaway, 
\begin{equation}
\mathcal{F}\big|_{\textrm{runaway }}=a^{4}+\frac{8}{a^{4}b^{6}},\qquad\mathcal{D}\big|_{\textrm{runaway }}=\frac{1}{8}\left(a^{2}-b^{2}-\frac{4}{a^{8}b^{4}}\right)^{2}.
\end{equation}
We ssume that $a\ll b^{-2/5}$ and so the term $a^{2}$ in $\mathcal{D}$
can be neglected, this will be justified by self-consistency.  In order to incorporate $\Delta$ dependence, we rescale
the fields (again):
\begin{equation}
b\rightarrow b\Delta^{-1/7},\qquad a\rightarrow a\Delta^{3/28}.\label{eq:abScaling}
\end{equation}
 Under this scaling, the full scalar potential is 
\begin{equation}
V=\Lambda^{4}\lambda^{10/7}\Delta^{3/7}\left[a^{4}+\frac{8}{a^{4}b^{6}}+\frac{1}{8}\left(b^{2}+\frac{4}{a^{8}b^{4}}\right)^{2}\right].
\end{equation}
The vacuum energy is $E\approx2.7\,\Lambda^{4}\lambda^{10/7}\Delta^{3/7}$
and the VEVs are given by 
\begin{equation}
\left\langle b\right\rangle \approx1.6\,\left(\lambda\,\Delta\right)^{-1/7}\Lambda,\qquad\left\langle a\right\rangle \approx1.0\,\left(\lambda/\Delta^{3/4}\right)^{-1/7}\Lambda\label{eq:g<h Vacuum Structure}
\end{equation}
with $\lambda$, $\Delta$, and $\Lambda$ reintroduced. This vacuum can be trusted as long as $\lambda^{5/3}\ll g_{2}\ll\lambda$. For $\Delta\sim1$, a numerical calculation shows a smooth interpolation between the two regimes, presented in figure \ref{abVacuum}. In the window $\lambda^{4}\lesssim g_{2}\lesssim\lambda^{5/3}$, quantum
corrections are not under control and the vacuum structure is unknown.
\newpage

As expected by the theorem proved in \cite{ZoharThomasSudanoDterms}, the ratio $\Delta\mathcal{D}/\mathcal{F}$ is always bounded. It decreases monotonically as a function of $\Delta$:
\begin{equation}
\frac{\Delta\mathcal{D}}{\mathcal{F}}\sim\begin{cases}
3/4 \quad & \Delta\rightarrow 0\\
\Delta^{-1} \quad& \Delta\rightarrow \infty.
\end{cases}
\end{equation}

\section{The 3-2 model with extra matter\label{sec:Adding-Extra-Matter}}

We now equip the 3-2 Model with two massive vector-like pairs of $SU(3)_{\textrm{C}}$,
$\Psi^{i}$,$\bar{\Psi}_{i}$, in order for a weakly coupled dual theory to exist in some region of parameter space. We restrict to the limit $\Delta\gg1$.\footnote{The opposite limit was studied in \cite{ZoharThomasSudanoDterms}.}

For very large masses, the vacuum structure does not know about the extra matter, and the low energy theory is essentially the same one previously discussed.
However, the effective strong coupling scale is higher since the $\Psi$'s
have been integrated out at the mass scale $m$; the scale matching is given by $\Lambda_{\mathrm{eff}}^{7}=\Lambda^{5}m^{2}$. The vacuum is located at
\begin{equation}
a\sim b\sim L_{2}\sim\left({\delta}/{\lambda}\right)^{1/7}\Lambda\label{eq:VacuumScalingLargeFields}
\end{equation}
where we defined $\delta\equiv\left(m/ \Lambda \right)^{2}$.
This analysis is rigorous as long as $\delta\gg1$. 

When $m$ is lowered beneath $\Lambda$, there is no longer a Wilsonian sense to integrate out the $\Psi$'s. Using holomorphy, we conclude that the only possible superpotential is
\begin{equation}
\mathcal{W}=\frac{2\Lambda^{5}m^{2}}{\mathrm{det}\left(\bar{Q}Q\right)}+\lambda \bar{Q}^{1}QL+m\mathrm{Tr}\left(\Psi\bar{\Psi}\right).
\label{HolomorphicSuperpotential}
\end{equation}
Since there are no mixed terms involving both $\Psi$'s and $Q,\bar{Q},L$,
the vacuum remains at (\ref{eq:VacuumScalingLargeFields}) and can
be trusted for $\delta<1$ in the regime $\delta\gg\lambda$ where it is controllable.\footnote{The limits of (\ref{HolomorphicSuperpotential}) should be taken with $\Lambda_{\mathrm{eff}}$ fixed.}

As the $\Psi$ mass is further reduced, the vacuum moves towards strong
coupling and a different analysis is required. If a vacuum exists
in the small fields regime, the description of the low energy physics
is given by the Seiberg dual magnetic theory. In the limit
$\Lambda_{3}\gg\Lambda_{2}$, only the
$SU(3)_{\textrm{C}}$ sector is dualized. The magnetic theory consists of
4 pairs of quarks $q^{i}$ and antiquarks
$\bar{q}_{j}$, $4\times4$ mesons $M_{i}^{j}$ and the doublet $L$. It is convenient to decompose the mesons and quarks as
\begin{equation}
M=\left(\begin{array}{cc}
\Phi_{i}^{j} & \Upsilon{}_{i}^{j}\\
\tilde{\Upsilon}_{A}^{j} & Z_{A}^{j}
\end{array}\right)=\frac{1}{\Lambda}\left(\begin{array}{cc}
\Psi_{i}\bar{\Psi}^{j} & \Psi_{i}\bar{Q}^{j}\\
Q_{A}\bar{\Psi}^{j} & Q_{A}\bar{Q}^{j}
\end{array}\right),\qquad
q=(x,\sigma^{A}),\qquad\bar{q}=(\bar{x},\bar{\sigma}_{i}).
\end{equation}
The magnetic gauge group consists only of the untouched $SU(2)_{\textrm{L}}$; $\tilde{\Upsilon}^{j}$, $Z^{j}$,  $\sigma$ and $L$  transform in their fundamental representation. $\sigma$ is reminiscent of the baryon $\frac{1}{2}\Psi_{[a}^{[i}\Psi_{b]}^{j}Q_{A}^{k]}$.

The superpotential is 
\begin{equation}
\mathcal{W}=q^{i}M_{i}^{j}\bar{q_{j}}-\mu^{2}\mathrm{Tr}\Phi+\lambda\Lambda Z^{1}L
\end{equation}
with $\mu^{2}\sim m\Lambda=\sqrt{\delta}\Lambda^{2}$.\footnote{ For the sake of clarity, the magnetic Yukawa coupling and other numeric
factors of $\mathcal{O}(1)$ are omitted.} 

A significant shortcut can now be taken. Consider a different theory with the
same matter content but with $\lambda$ sent to zero and $SU(2)_{\textrm{L}}$ a global symmetry. 
This type of theory has been studied in \cite{FrankoUranga}, the relevant
details are presented here for completeness. 

The F-term for $\Phi_{i}^{j}$
breaks SUSY by the rank condition. Minimization of the tree
level potential is achieved by saturating one component of $x$ 
\begin{equation}
\left\langle x\right\rangle =\left\langle \bar{x}\right\rangle ^{T}=(\mu,0)\label{eq:RankConditionVEV}
\end{equation}
and the vacuum energy is $\mu^{4}\sim\delta\Lambda^{4}$. The mass
spectrum of the fluctuations is identified by expanding the superpotential
around (\ref{eq:RankConditionVEV}). At tree level, some fields get a
mass of order $\mu$. \newline The one loop Coleman-Weinberg potential \cite{ColemanWeinberg} stabilizes all of the fields that remained massless except $Z_{A}^{j}$. Henceforth, it will be diagonalized using $SU(2)_{\textrm{L}}\times SU(2)_{\textrm{R}}$.\footnote{As in the electric description, the $SU(2)_{\textrm{R}}$ flavor symmetry is broken by turning on $\lambda$. Future constraints on $|\lambda|$ ensure that this symmetry breaking is small.} 

At two loops, this pseudo modulus is destabilized \cite{DSBSQCD,SQCDMassless,GoodBadUgly}.
For $Z_{A}^{i}\gg\mu$,\begin{equation}
V_{\mathrm{eff}}^{(2)}\sim-\frac{1}{\left(16\pi^{2}\right)^{2}}\mu^{4}\left[\textrm{log}^{2}\left({\left|Z_{1}^{1}\right|^{2}}/{\mu^{2}}\right)+\textrm{log}^{2}\left({\left|Z_{2}^{2}\right |^{2}}/{\mu^{2}}\right)\right]
\end{equation}
is a good approximation to the effective potential.

As before, $\lambda$ is turned on in attempt to stabilize the runaway, leading to \begin{equation}
V_{\mathrm{tree}}\supset\left(\lambda\Lambda\right)^{2}\left|Z_{A}^{1}\right|+\left|\lambda\Lambda L^{A}+\sigma^{A}\bar{\sigma}_{1}\right|^{2}=\left(\lambda\Lambda\right)^{2}\left(\left|Z_{A}^{1}\right|^{2}+\left|L^{A}\right|^{2}\right)
\end{equation}
where the last equality holds since $\sigma$, $\bar{\sigma}$ are
stabilized at the origin. As in the electric description, the tree
level interaction does not lift all runaway directions unless $SU(2)_{\textrm{L}}$ is weakly gauged. In the limit $\Delta\gg1$, the vacuum sits at the minimum of $\left(V_{\mathrm{tree}}+V_{\mathrm{eff}}^{(2)}\right)\Big|_{\mathcal{D}=0}$
where 
\begin{equation}
\mathcal{D}\equiv\frac{1}{8}\left[4\left|L_{1}L_{2}\right|^{2}+\left(\left|Z_{1}^{1}\right|^{2}-\left|Z_{2}^{2}\right|^{2}+\left|L_{1}\right|^{2}-\left|L_{2}\right|^{2}\right)^{2}\right].
\end{equation}
The general solution to $\mathcal{D}=0$ has two possible parameterizations:
\begin{equation}
L=\Big(0,\sqrt{\left|Z_{1}^{1}\right|^{2}-\left|Z_{2}^{2}\right|^{2}}\Big),\qquad 
L=\Big(\sqrt{\left|Z_{2}^{2}\right|^{2}-\left|Z_{1}^{1}\right|^{2}},0\Big).
\end{equation}
Since there is no gauge symmetry left to connect the two solutions, $L$ must be stabilized at the origin for the magnetic dual to have the same global symmetry as the underlying theory. Indeed, this will turn out to be the case. 

It is useful to absorb all constants by a redefinition of $Z$: 
\begin{equation}
Z_{i}^{i}\equiv\mu\sqrt{a}\zeta_{(i)}
\end{equation}
where $a\equiv 2\left(16\pi^{2}\right)^{-2}\lambda^{-2}\delta^{1/2}$. Note that $Z_{A}^{i}\gg\mu$ means $a\gg1$. Using these variables, the effective potential is 
\begin{equation}
V_{\mathrm{eff}}=\frac{\mu^{4}}{\left(16\pi^{2}\right)^{2}}\left[2\zeta_{(1)}^{2}+2\left|\zeta_{(1)}^{2}-\zeta_{(2)}^{2}\right|-\textrm{log}^{2}\left(a\;\zeta_{(1)}^{2}\right)-\textrm{log}^{2}\left(a\;\zeta_{(2)}^{2}\right)\right].\label{eq:LargeZPotential}
\end{equation}
For a given $\zeta_{(1)}$, a minimum is found for $\zeta_{(2)}=\zeta_{(1)}\equiv\zeta$. As promised, $L$ is stabilized at the origin.  We expect this to hold to all orders in perturbation theory.

Expanding around large values of $\zeta$, we find
\begin{equation}
\left\langle Z_{A}^{i}\right\rangle =\mu\sqrt{2\, a\,\textrm{log}\left[a\right]}\sim\delta^{\frac{1}{2}}\lambda^{-1}\Lambda\frac{1}{16\pi^{2}}.
\end{equation}
For the magnetic description to be reliable, $\left\langle Z_{A}^{i}\right\rangle \ll\Lambda$, or equivalently $\delta\ll\lambda^{2}$ must hold.

One might question whether this analysis is at all valid as the loop
computations were not expanded around the true tree level vacuum,
since the {}``Yukawa'' coupling $\lambda$ was introduced after
the fact. In this respect, self-consistency requires that the mass
given to $Z_{A}^{1}$ and $L_{A}$ by turning on $\lambda$ be much
smaller than all other tree level masses: $\lambda\Lambda\ll\mu$.
Luckily, this condition is equivalent to $a\gg1$.

The weak gauging of $SU(2)_{\textrm{L}}$ poses another question: might the two
loop effective potential itself be altered by a contribution from
gauge fields? The reason this does not happen is that all fields charged
under $SU(2)_{\textrm{L}}$ have supersymmetric masses at tree level and so the
gauge spectrum is supersymmetric at one loop and therefore cannot
contribute to the effective potential at two loops.

The electric and magnetic descriptions are seen to be valid in non-overlapping regions of parameter space, $\lambda\ll\delta$ and $\delta\ll\lambda^{2}$
respectively. As advocated, this is a consistency check of the duality. Moreover, the fields which play a nontrivial role in the
vacuum structure of the magnetic dual are closely related to the gauge
invariants of the electric theory even though the source of the runaway
differs significantly between the two descriptions.  

\section{Summary}

$\bullet$ The vacuum structure of the 3-2 model was studied for generic values of the gauged $SU(2)_{\textrm{L}}$ coupling $g_{2}$ and the {}``Yukawa'' coupling $\lambda$. 
A controllable vacuum is found in the regime $g_{2} \lesssim\lambda$ even when the non-perturbative interactions are generated by an instanton.\newline
$\bullet$  Two vector-like pairs of $SU(3)_{\textrm{C}}$ with mass $m=\sqrt{\delta}\Lambda$ were added to the theory, and the vacuum was found in the limit $g_{2}\gg\lambda$  in two regions of parameter space. For $\lambda\ll\delta$ the vacuum is at large values of the microscopic fields whereas for $\lambda^{2}\gg\delta$ these fields confine and the vacuum is described by the magnetic dual theory. In both regions, the combined efforts of a tree level superpotential and the weak gauging of $SU(2)_{\textrm{L}}$ stabilize a runaway
potential. The fact that the two regions of validity do not overlap is a consistency check of the duality. In the case studied here, it depends nontrivially on non-holomorphic data.

For $\lambda^{2}\lesssim\delta\lesssim\lambda$, the vacuum is nowhere to be found as both descriptions break down; what happens here remains an open question.

\acknowledgments

I would like to thank Roberto Auzzi, Shmuel Elitzur, Bjarke Gudnason, Amihay Hanany, Kenneth Intriligator, Eliezer Rabinovici and especially Thomas Dumitrescu for interesting conversations and useful comments.
I am grateful for illuminating discussions with Zohar Komargodski
and Amit Giveon, who also suggested this project. 
This work was supported
in part by the BSF \textendash{} American-Israel Bi-National Science
Foundation, and by a center of excellence supported by the Israel
Science Foundation~(grant~1665/10). 

\newpage

\bibliographystyle{unsrt}

\end{document}